\documentclass[aps,twocolumn,superscriptaddress,showpacs]{revtex4}

\usepackage{amsmath}
\usepackage{graphicx}
\usepackage{epsfig}             

\begin{document}


\title{Spatial state Stokes-operator squeezing and entanglement for optical beams}

\author{M.~T.~L.~Hsu}
\affiliation{ARC COE for Quantum-Atom Optics, Australian National University, Canberra, ACT 0200, Australia.}
\affiliation{Present address: Department of Chemistry, Stanford University, Palo Alto, CA 94305, U.S.A.}

\author{W.~P.~Bowen}
\affiliation{School of Physical Sciences, University of Queensland,
Brisbane, QLD 4072, Australia.}

\author{P.~K.~Lam}
\affiliation{ARC COE for Quantum-Atom Optics, Australian National University, Canberra, ACT 0200, Australia.}

\date{\today}

\begin{abstract}
The transverse spatial attributes of an optical beam can be decomposed into the position, momentum and orbital angular momentum observables. The position and momentum of a beam is directly related to the quadrature amplitudes, whilst the orbital angular momentum is related to the polarization and spin variables. In this paper, we study the quantum properties of these spatial variables, using a representation in the Stokes-operator basis. We propose a spatial detection scheme to measure all three spatial variables and consequently,  propose a scheme for the generation of spatial Stokes operator squeezing and entanglement.
\end{abstract}

\pacs{42.50, 42.30}

\maketitle

\section{Introduction}

Squeezed and entangled bright optical fields are essential resources
in the continuous variable quantum optics and quantum information
communities \cite{CerfLeuchsBook}. To date, the vast majority of
research has been focused on fields which exhibit non-classical
features on their amplitude and phase quadratures. Recently,
however, a number of papers have been published on squeezing and
entanglement of other field variables, and in particular the
polarization \cite{PolarizationSQZ} and spatial structure
\cite{SpatialSQZ, HsuJOB, HsuPRA05, Wagner}. These variables can be directly related to momentum components of the field. The quadrature amplitudes are associated with the {\it transverse linear momentum} of the field, whilst the polarization states and spatial states considered to date are respectively associated to the {\it spin angular momentum} and the {\it transverse angular momentum}. We immediately see why interaction of a polarization squeezed field with an atomic ensemble can yield atomic spin
squeezing \cite{Hald}, and why spatially entangled light can be used to test the EPR paradox with position and momentum variables \cite{HsuPRA05} as was originally proposed by Einstein, Podolsky and Rosen \cite{EPR}.

The correspondence of non-classical polarization and spatial states
with non-classical momentum states immediately raises the question
of whether non-classical {\it orbital angular momentum} states can
also be generated. Such states have been under investigation for
some time in the discrete variable regime.  Techniques have been
established to detect the orbital angular momentum properties of
single photons \cite{leach, leach04, langford}; and discrete-variable
multi-dimensional entanglement between orbital angular momentum
states has been proposed \cite{arnaut,franke-arnold} and
demonstrated \cite{mair}. However, to date, non-classical orbital
angular momentum states have not been investigated for the continuous variable regime, that is relevant in many applications \cite{allen, simpson, arlt, barnett, leach04B, malyutin, beijersbergen}. We investigate these continuous variable orbital angular momentum quantum states,
proposing techniques for both generation and detection. Many
applications have been proposed for optical orbital angular momentum
in the classical domain, ranging from the generation of
counter-rotating superpositions in Bose-Einstein condensates
\cite{kapale, isoshima}, and transfer of orbital angular momentum
from an atomic ensemble to a light field \cite{akamatsu}, to optical
angular momentum transfer to trapped particles \cite{friese, he,
paterson}. It should be noted that light with orbital angular momentum
has also been applied to achieve super-resolution imaging of molecules and proteins in biological systems, via techniques such as {\it stimulated emission depletion} (STED) \cite{hell}. Non-classical orbital angular momentum states offer the prospect to improve these processes, and to generate non-classical orbital angular momentum states in macroscopic physical systems.

Ref.~\cite{HsuPRA05} proposed EPR entanglement of the
position-momentum observables of optical fields, and this has
recently been experimentally observed \cite{Wagner}. In these references,
entanglement was present only for one transverse axis of the beam,
and analysis only required consideration of the TEM$_{00}$ and TEM$_{01}$ modes. One of the strengths of entanglement in the spatial domain is
access to higher dimensional spaces. Here we extend the work of Refs.~\cite{HsuPRA05, Wagner} to include both transverse beam axes, in a formalism easily extended to higher order TEM modes. In the process, orbital angular momentum is introduced as an entanglement variable in addition to beam position and momentum. These quantum variables can be represented on a spatial version of the Poincar\'e sphere commonly used for analysis of polarization quantum states. We propose a scheme to measure their signal and noise properties, via a spatial Stokes detection scheme;
and finally introduce a scheme to generate spatial Stokes-operator
squeezing and entanglement.

\section{Spatial Stokes operators}

\begin{figure}[!ht]
\begin{center}
\includegraphics[width=8cm]{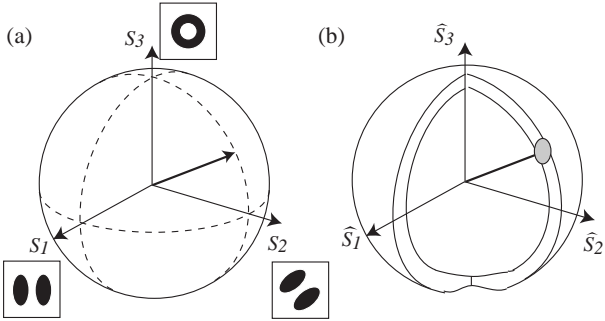}
\caption{(a) Poincar\'e sphere representation based on the spatial modes TEM$_{10}$ and TEM$_{01}$. $S_1$ is the Stokes variable for the $u_{01}(\boldsymbol{r})$  and $u_{10}(\boldsymbol{r})$ modes, $S_2$ is the Stokes variable for the $u_{01}^{45^{\circ}}(\boldsymbol{r})$ and $u_{10}^{45^{\circ}}(\boldsymbol{r})$ modes, and $S_3$ is the Stokes variable for the $u_{01}^{+1}(\boldsymbol{r})$ and $u_{01}^{-1}(\boldsymbol{r})$  modes. (b) Poincar\'e sphere representation for the spatial Stokes operators. The shaded area indicates the quantum noise associated with the mean amplitude of the Stokes operator.}
\label{poincare}
\end{center}
\end{figure}

Spatial quantum states exist in an infinite dimensional Hilbert space, which may be conveniently expanded in a basis of TEM$_{pq}$ modes. After such an expansion the positive frequency part of the electric field operator, $\hat{\mathcal{E}}^{+}(\boldsymbol{r})$, can be written explicitly in terms of the double-subscript sum
\begin{equation}
\hat{\mathcal{E}}^{+} (\boldsymbol{r}) = i \sqrt{\frac{\hbar \omega}{2 \epsilon_{0} V}} \sum_{p,q=0}^{\infty} \hat{a}_{pq} u_{pq} (\boldsymbol{r})
\end{equation}
where $\hat{a}_{pq}$ and $u_{pq} (\boldsymbol{r})$ are the photon annihilation operator and the normalized transverse beam amplitude function, respectively, associated with the TEM$_{pq}$ mode. The photon annihilation operator $\hat{a}_{pq}$ can be formally defined with the projection of the transverse beam amplitude function with the positive frequency part of the electric field operator, given by
\begin{equation}
\hat{a}_{pq} = \iint_{\infty}^{-\infty} dx dy \hat{\mathcal{E}}^{+} (x,y) u_{pq} (x,y).
\end{equation}
Within this basis, it is natural to consider pairs of modes with amplitude function that correspond to a physical $\pi/2$ rotation in the transverse plane. That is, modes of the form TEM$_{pq}$ and TEM$_{qp}$, with $p$ and $q$ interchanged ($p \neq q$). Pairs of this form are naturally described  by a set of spatial Stokes operators $\hat S_i^{(p,q)}$ where $i \in \{0,1,2,3\}$, that can be represented by a Poincar\'e sphere \cite{padgett} in direct analogy to polarization Stokes operators \cite{fano-stokes}, as shown in Fig.~\ref{poincare}.

Using the definition of the classical Stokes parameters \cite{padgett}, corresponding quantum mechanical spatial Stokes operators are defined as
\begin{eqnarray} \label{stokesoperators}
\hat{S}_{0} & = &\hat{a}_{pq}^{\dagger} \hat{a}_{pq} + \hat{a}_{qp}^{\dagger} \hat{a}_{qp} \nonumber\\
\hat{S}_{1} & = &\hat{a}_{pq}^{\dagger} \hat{a}_{pq} - \hat{a}_{qp}^{\dagger} \hat{a}_{qp} \nonumber\\
\hat{S}_{2} &= &\hat{a}_{pq}^{\dagger} \hat{a}_{qp}e^{i\theta} + \hat{a}_{qp}^{\dagger} \hat{a}_{pq}e^{-i\theta} \nonumber\\
\hat{S}_{3} &= & i\hat{a}_{qp}^{\dagger} \hat{a}_{pq}e^{-i\theta} - i\hat{a}_{pq}^{\dagger} \hat{a}_{qp}e^{i\theta}\nonumber\\
\end{eqnarray}
where $\theta$ is the phase difference between the TEM$_{pq}$ and TEM$_{qp}$ modes.  As is the case with polarization, the spatial Stokes operators obey commutation relations. By using the commutation relations of the photon annihilation and creation operators, $[\hat{a}_{mn}, \hat{a}_{jk} ] = \delta_{mj} \delta_{nk}$, the Stokes-operator commutation relations can be found to be
\begin{equation} \label{commrel}
\left[ \hat{S}_{i}, \hat{S}_{j} \right]  =  2i \epsilon_{ijk} \hat{S}_{k}
\end{equation}
where $\epsilon_{ijk}$ is the Levi-Civita tensor.

Each Stokes operator characterizes a different spatial property of the field. $\hat{S}_{0}$ represents the beam intensity, whilst the Stokes vector $(\hat{S}_{1}, \hat{S}_{2}, \hat{S}_{3})$ characterizes its transverse momentum. $\hat{S}_{1}$ and $\hat{S}_{2}$ respectively quantify the relative proportions of horizontal to vertical transverse momentum, and $45^\circ$ to $-45^\circ$ diagonal transverse momentum; whilst $\hat{S}_{3}$ weights left to right orbital angular momentum.

\begin{figure}[!ht]
\begin{center}
\includegraphics[width=8cm]{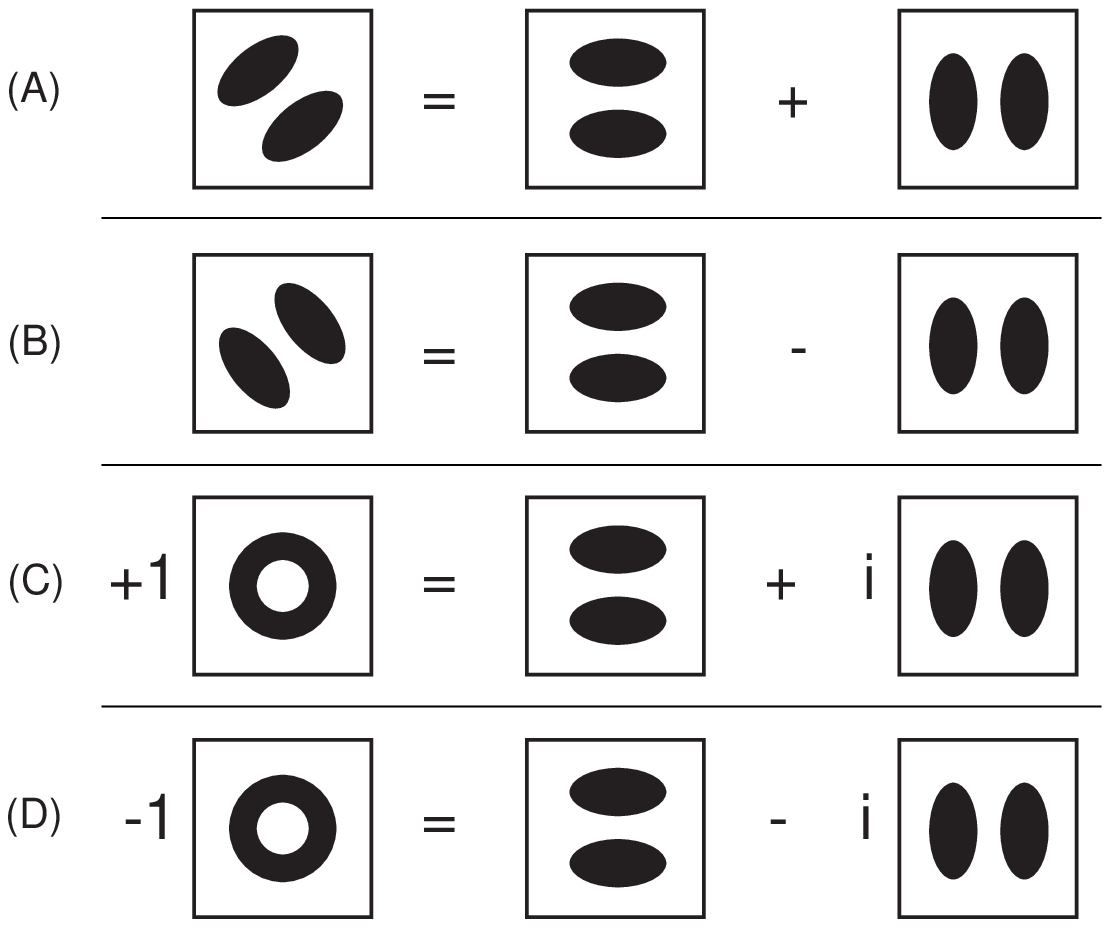}
\caption{Intensity distribution for the modes given by (A) $u_{01}^{45^{\circ}}  (\boldsymbol{r})= u_{01} (\boldsymbol{r}) + u_{10} (\boldsymbol{r})$, (B) $u_{10}^{45^{\circ}}  (\boldsymbol{r})= u_{01} (\boldsymbol{r}) - u_{10} (\boldsymbol{r})$, (C) $u^{+1}_{01} (\boldsymbol{r})= u_{01}(\boldsymbol{r}) + i u_{10}(\boldsymbol{r})$,  and (D) $u^{-1}_{01} (\boldsymbol{r})= u_{01}(\boldsymbol{r}) - i u_{10}(\boldsymbol{r})$.}
\label{lghgmode}
\end{center}
\end{figure}

The spatial Poincar\'e sphere can be fully spanned by spatially overlapping two orthogonal TEM$_{pq}$ and TEM$_{qp}$ modes. The diagonal modes, for example, are given by $u_{pq}^{45^{\circ}} (\boldsymbol{r}) = u_{qp} (\boldsymbol{r}) - u_{pq} (\boldsymbol{r})$ and $u_{qp}^{45^{\circ}}  (\boldsymbol{r})= u_{qp} (\boldsymbol{r}) + u_{pq} (\boldsymbol{r})$, and respectively yield Stokes vectors oriented along the negative and positive $\hat S_2$ axis of the Poincar\'e sphere. The Laguerre-Gauss modes LG$_{0q}$ with orbital angular momentum $q$ and $-q$, are given by $u^{+l}_{0q} (\boldsymbol{r})= u_{0q}(\boldsymbol{r}) +i u_{q0}(\boldsymbol{r})$ and $u^{-l}_{0q}(\boldsymbol{r}) = u_{0q}(\boldsymbol{r}) -i u_{q0}(\boldsymbol{r})$, yielding Stokes vectors oriented along the $\hat S_3$ axis. For example, the intensity distributions for the diagonal and Laguerre-Gauss modes, generated from the combinations of TEM$_{10}$ and TEM$_{01}$ modes, are shown in Fig.~\ref{lghgmode}.

Since the diagonal and orbital modes can be generated purely from two orthogonal TEM$_{pq}$ and TEM$_{qp}$ modes, only by changing the phase between the modes, these sets of modes obey the SU(2) group properties. At this point, we restrict our analysis to the TEM$_{10}$ and TEM$_{01}$ modes, for illustrative purposes. In principle, our analysis is valid for all orthogonal TEM$_{pq}$ and TEM$_{qp}$ modes. 

\section{Spatial Stokes detection}

Now that the spatial Stokes operators have been defined, we consider a system for their detection. Analogous to polarization Stokes operator detection, the spatial Stokes detection requires two separate photodiodes, a modal phase shifter, and a mode separator.

\begin{figure}[!ht]
\begin{center}
\includegraphics[width=8cm]{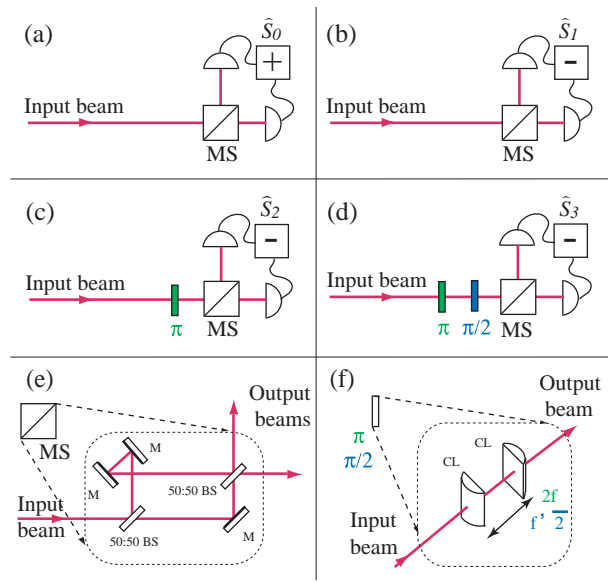}
\caption{Measurements of (a) $\hat{S}_{0}$, (b) $\hat{S}_{1}$, (c) $\hat{S}_{2}$ and (d) $\hat{S}_{3}$. (e) An example of a mode separator (MS) is an asymmetric Mach-Zehnder interferometer. (f) The $\pi$ and $\pi/2$ modal phase shifters could be constructed using a pair of cylindrical lenses with lens separation given by $2f$ and $\sqrt{2}f$, respectively. $f$ is the focal length of the cylindrical lenses, M is a mirror, 50:50 BS is a symmetric non-polarizing beam-spliter.}
\label{stokesmeasurement}
\end{center}
\end{figure}

A modal phase shifter is a device that introduces a relative phase shift between the two orthogonal TEM$_{10}$ and TEM$_{01}$ modes. The modal phase shifter can be constructed using a pair of cylindrical lenses with variable lens separation, as shown in Fig.~\ref{stokesmeasurement}~(f). The pairing of cylindrical lenses introduces an astigmatic Gouy phase shift between the TEM$_{10}$ and TEM$_{01}$ modes. In order to introduce $\pi$ and $\pi/2$ modal phase shift, the lens separation is given by $2f$ and $\sqrt{2}f$, respectively, where $f$ is the focal length of the cylindrical lens. Ref.~\cite{beijersbergen} contains a detailed analysis of the $\pi$ and $\pi/2$ modal phase shifters.

A mode separator (MS) is a device that separates the two orthogonal TEM$_{10}$ and TEM$_{01}$ modes and can be constructed using an asymmetric Mach-Zehnder interferometer \cite{delaubert}, as shown in
Fig.~\ref{stokesmeasurement}~(e). Due to the odd and even numbers of
reflections in each interferometer arm, different interference
conditions for the TEM$_{01}$ and TEM$_{10}$ modes are present. The
resulting outputs from the interferometer is a separation of even
and odd (defined along one spatial axis) spatial modes \cite{delaubert}.

Measurements of the total signal and noise, which correspond to a
measurement of $\hat{S}_{0}$, are given by the sum of the photocurrents as shown in Fig.~\ref{stokesmeasurement}~(a). Measurement of $\hat{S}_{1}$ involves taking the subtraction between the photocurrent outputs corresponding to mode components TEM$_{10}$ and TEM$_{01}$,  as shown in Fig.~\ref{stokesmeasurement}~(b). Measurement of $\hat{S}_2$ involves subtraction of the diagonal mode components and is obtained by phase shifting one mode by $\pi$ with respect to the other and then taking the subtraction of the photocurrent signals, given in Fig.~\ref{stokesmeasurement}~(c). $\hat{S}_3$ is measured by decomposing the Laguerre-Gauss mode into its TEM$_{10}$ and TEM$_{01}$ modes using $\pi$ and $\pi/2$ modal phase shifters, as shown in Fig.~\ref{stokesmeasurement}~(d).

The photon annihilation operators in Eq.~(\ref{stokesoperators}) can
be written in the form $\hat{a}_{pq} = \alpha_{pq} + \delta
\hat{a}_{pq}$, where $\alpha_{pq}$ describes the mean amplitude part
and $\delta \hat{a}_{pq}$ is the quantum noise operator. Using the
linearized formalism, where second order terms in the fluctuation
operator are neglected (i.e. $|\alpha_{pq} |^2 \gg | \langle \delta
\hat{a}_{pq}^2 \rangle |$), the mean amplitudes of the Stokes
operators in Eq.~(\ref{stokesoperators}), in terms of the TEM$_{10}$ and TEM$_{01}$ modes, are therefore given by
\begin{eqnarray} \label{stokesopsdef}
\langle \hat{S}_{0} \rangle & = & \alpha_{\rm 10} ^2 +  \alpha_{\rm 01} ^2 = N \nonumber\\
\langle \hat{S}_{1} \rangle & = & \alpha_{\rm 10} ^2 -  \alpha_{\rm 01} ^2  \nonumber\\
\langle \hat{S}_{2} \rangle & = & 2 \alpha_{\rm 10} \alpha_{\rm 01}  \cos \theta \nonumber\\
\langle \hat{S}_{3} \rangle & = & 2 \alpha_{\rm 10} \alpha_{\rm 01}  \sin \theta
\end{eqnarray}
where $\alpha_{01}$ and $\alpha_{10}$ are the mean amplitude terms corresponding to modes $u_{01} (\boldsymbol{r})$ and $u_{10}(\boldsymbol{r})$, respectively. $\langle \hat{S}_{0} \rangle = N$ is the total mean number of photons, $\langle \hat{S}_{1} \rangle$ is the difference in the mean number of photons in the $u_{10} (\boldsymbol{r})$ and $u_{01}(\boldsymbol{r})$ modes, $\langle \hat{S}_{2} \rangle$ is the difference in the mean number of photons in the $u_{01}^{45^{\circ}}(\boldsymbol{r})$ and $u_{10}^{45^{\circ}}(\boldsymbol{r})$ modes and $\langle \hat{S}_{3} \rangle$ is the difference in the mean number of photons in the $u_{01}^{+1}(\boldsymbol{r})$ and $u_{01}^{-1}(\boldsymbol{r})$ modes.

The Stokes operators of Eq.~(\ref{stokesoperators}) can be expanded
in terms of quadrature operators with the general form
$\hat{X}_{a_{pq}}^{\phi} = e^{-i\phi}\delta \hat{a}_{pq} + e^{i\phi}
\delta \hat{a}_{pq}^{\dagger}$, and their variances are then given
in general by
\begin{eqnarray} \label{stokesvar}
\langle (\delta \hat{S}_{0})^2 \rangle & = & \alpha_{10}^2  \langle (\delta \hat{X}_{a_{10}}^+)^2 \rangle +  \alpha_{01}^2 \langle (\delta \hat{X}_{a_{01}}^+)^2 \rangle \nonumber\\
 & & + 2  \alpha_{\rm 10} \alpha_{01} \langle \delta \hat{X}_{a_{10}}^+ \delta \hat{X}_{a_{01}}^+ \rangle \nonumber\\
\langle (\delta \hat{S}_{1})^2 \rangle  & = & \alpha_{\rm 10}^2  \langle (\delta \hat{X}_{a_{\rm 10}}^+)^2 \rangle +  \alpha_{\rm 01}^2 \langle (\delta \hat{X}_{a_{\rm 01}}^+)^2 \rangle \nonumber\\
 & & - 2  \alpha_{\rm 10}  \alpha_{\rm 01} \langle \delta \hat{X}_{a_{\rm 10}}^+ \delta \hat{X}_{a_{\rm 01}}^+ \rangle  \nonumber\\
\langle (\delta \hat{S}_{2})^2 \rangle  & = & \alpha_{\rm 10}^2  \langle (\delta \hat{X}_{a_{\rm 01}}^{-\theta})^2 \rangle +  \alpha_{\rm 01}^2 \langle (\delta \hat{X}_{a_{\rm 10}}^{\theta})^2 \rangle \nonumber\\
 & &  + 2  \alpha_{\rm 10}  \alpha_{\rm 01} \langle \delta \hat{X}_{a_{\rm 01}}^{-\theta} \delta \hat{X}_{a_{\rm 10}}^{\theta} \rangle \nonumber\\
\langle (\delta \hat{S}_{3})^2 \rangle  & = & \alpha_{\rm 10}^2  \langle (\delta \hat{X}_{a_{\rm 01}}^{-\theta+\frac{\pi}{2}})^2 \rangle + \alpha_{\rm 01}^2 \langle (\delta \hat{X}_{a_{\rm 10}}^{\theta-\frac{\pi}{2}})^2 \rangle \nonumber\\
 & &  + 2  \alpha_{\rm 10}  \alpha_{\rm 01} \langle \delta \hat{X}_{a_{\rm 01}}^{-\theta+\frac{\pi}{2}} \delta \hat{X}_{a_{\rm 10}}^{\theta-\frac{\pi}{2}} \rangle \nonumber\\
\end{eqnarray}
where $\hat{X}_{a_{pq}}^{+} = \hat{X}_{a_{pq}}^{\phi=0}$ and
$\hat{X}_{a_{pq}}^{-} = \hat{X}_{a_{pq}}^{\phi=\pi/2}$ are respectively the amplitude and phase quadrature operators of the TEM$_{pq}$ mode.

Non-classical optical orbital angular momentum states can be constructed by spatially overlapping a set of orthogonal non-classical TEM$_{pq}$ fields, in analogy to the work of Refs.~\cite{HsuPRA05, Wagner} for transverse spatial entanglement. In such a scenario, no correlations should exist between the quadratures of the different input fields, such that $\langle \delta \hat{X}_{a_{10}}^{\phi_1} \delta\hat{X}_{a_{01}}^{\phi_2} \rangle = 0$, $\forall \{ \phi_1,\phi_2 \}$. Making this assumption, which we will adopt henceforth, the variances of the Stokes operators are simplified to
\begin{eqnarray} \label{stokesops}
\langle (\delta \hat{S}_{0})^2 \rangle & = & \alpha_{10}^2  \langle (\delta \hat{X}_{a_{10}}^+)^2 \rangle +  \alpha_{01}^2 \langle (\delta \hat{X}_{a_{01}}^+)^2 \rangle\nonumber\\
\langle (\delta \hat{S}_{1})^2 \rangle  & = & \alpha_{\rm 10}^2  \langle (\delta \hat{X}_{a_{\rm 10}}^+)^2 \rangle +  \alpha_{\rm 01}^2 \langle (\delta \hat{X}_{a_{\rm 01}}^+)^2 \rangle  \nonumber\\
\langle (\delta \hat{S}_{2})^2 \rangle  & = & \alpha_{\rm 10}^2  \langle (\delta \hat{X}_{a_{\rm 01}}^{-\theta})^2 \rangle +  \alpha_{\rm 01}^2 \langle (\delta \hat{X}_{a_{\rm 10}}^{\theta})^2 \rangle   \nonumber\\
\langle (\delta \hat{S}_{3})^2 \rangle  & = & \alpha_{\rm 10}^2
\langle (\delta \hat{X}_{a_{\rm 01}}^{-\theta+\frac{\pi}{2}})^2
\rangle + \alpha_{\rm 01}^2 \langle (\delta \hat{X}_{a_{\rm
10}}^{\theta-\frac{\pi}{2}})^2 \rangle \nonumber.\\
\end{eqnarray}

\section{Spatial Stokes squeezing}

An examination of Eq.~(\ref{stokesops}) shows that simultaneous
squeezing of at least two spatial Stokes operators is possible,
using two quadrature squeezed input fields. Hence, a spatial Stokes squeezed state can be used to enhance relative measurements of momentum variables along multiple axes. This is best illustrated by considering a few examples.

Simultaneous squeezing of $\hat{S}_{0}$, $\hat{S}_{1}$ and $\hat{S}_{2}$ can be achieved through the in-phase ($\theta = 0$) spatial overlap of amplitude squeezed TEM$_{10}$ and TEM$_{01}$ modes (i.e. $ \langle (\delta \hat{X}_{a_{10}}^+)^2 \rangle < 1$ and $ \langle (\delta \hat{X}_{a_{01}}^+)^2 \rangle < 1$). This spatial squeezed state could be combined with a bright TEM$_{00}$ beam to allow enhanced measurements of the transverse momentum of the TEM$_{00}$ beam along any axis on the Poincar\'e sphere within the plane formed by $\hat{S}_1$ and $\hat{S}_2$. Simultaneous squeezing is exhibited along the horizontal/vertical axis and the diagonal/anti-diagonal axis.

Squeezing of $\hat{S}_{0}$, $\hat{S}_{1}$ and $\hat{S}_{3}$ can be achieved by overlapping amplitude squeezed TEM$_{10}$ and TEM$_{01}$ modes with phase difference $\theta = \pi/2$. The combination of this spatial squeezed beam with a bright TEM$_{00}$ beam would simultaneously enable sub-shot noise relative measurements of transverse momentum between the horizontal and vertical axes, as well as left- and right-handed orbital angular momentum of a TEM$_{00}$ beam \cite{HsuJOB}.

To achieve quantum enhanced measurements of transverse momentum between the diagonal and anti-diagonal axes, as well as between left- and right-handed orbital angular momentum of a TEM$_{00}$ beam, would require squeezing of $\hat{S}_{2}$ and $\hat{S}_{3}$. This can be achieved by overlapping quadrature squeezed TEM$_{10}$ and TEM$_{01}$ modes, at squeezed quadrature angle of $\pi/4$ (i.e. $\langle (\delta \hat{X}_{a_{10}}^{\frac{\pi}{4}})^2 \rangle < 1$ and
$ \langle (\delta \hat{X}_{a_{01}}^{\frac{\pi}{4}})^2 \rangle < 1$) and $\theta = \pi/4$.

Phase squeezed TEM$_{10}$ and TEM$_{01}$ modes (i.e. $ \langle
(\delta \hat{X}_{a_{10}}^-)^2 \rangle < 1$ and $ \langle (\delta
\hat{X}_{a_{01}}^-)^2 \rangle < 1$) will yield squeezing of $\langle
(\delta \hat{S}_2)^2 \rangle$ only.

\section{Spatial Stokes entanglement}

\begin{figure}[!ht]
\begin{center}
\includegraphics[width=8cm]{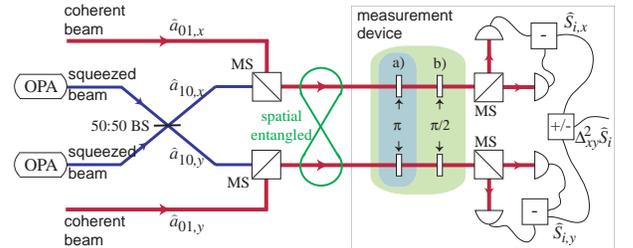}
\caption{Scheme to generate and characterize spatial Stokes entanglement. 50:50 BS is a symmetric non-polarizing beam-splitter, $\pi$ and $\pi/2$ are modal phase shifters.}
\label{stokesentg}
\end{center}
\end{figure}

Spatial Stokes entanglement can be generated as is shown in
Fig.~\ref{stokesentg}. Two squeezed TEM$_{10}$ beams, labeled respectively by the subscripts $x$ and $y$, are combined
with a $\pi/2$ relative phase shift on a 50:50 beam-splitter. The beams
after the beam-splitter, also in TEM$_{10}$ modes, exhibit the usual
quadrature entanglement. Two spatial mode combiners, each consisting
of a spatial mode separator as shown in Fig.~\ref{stokesmeasurement}
(e) in reverse, are then used to combine the quadrature entangled
beams with bright coherent TEM$_{01}$ beams, labelled here
$\hat{a}_{01,x}$ and $\hat{a}_{01,y}$, respectively. As we will show
here, the resulting output beams are entangled in the spatial Stokes
operator basis. Experimental verification of the spatial Stokes
entanglement can be performed using the detection scheme described
in the preceding sections.

To verify the existence of spatial Stokes entanglement we will use the generalized version of the Duan inseparability criterion \cite{duan} given in Ref.~\cite{Bowen}. This criterion provides a sufficient condition for entanglement and is given by
\begin{equation}
\Delta^2_{x\pm y} \hat{A} + \Delta^2_{x\pm y} \hat{B} < 2 | [ \delta \hat{A}, \delta \hat{B} ] |
\end{equation}
where $\hat{A}$ and $\hat{B}$ are two general observables, and $\Delta^2_{x\pm y} \hat{\mathcal{O}} = {\rm min} \langle ( \delta  \hat{\mathcal{O}}_{x} \pm \delta  \hat{\mathcal{O}}_y )^2 \rangle$. The degree of inseparability $\mathcal{I} (\hat{A}, \hat{B})$ is then given by \cite{Bowen}
\begin{equation}
\mathcal{I} (\hat{A}, \hat{B}) = \frac{ \Delta^2_{x\pm y} \hat{A} + \Delta^2_{x\pm y} \hat{B} }{ 2 | [ \delta \hat{A}, \delta \hat{B} ] | },
\end{equation}
where $\mathcal{I} (\hat{A}, \hat{B}) < 1$ indicates an inseparable state.

Assuming for simplicity and symmetry (i.e. both beams $x$ and $y$ are identically interchangeable) that $\alpha_{01,x} = \alpha_{01,y} = \alpha_ {01}$ and $\alpha_{10,x} = \alpha_{10,y} = \alpha_{10}$, and using Eqs.~(\ref{commrel}) and (\ref{stokesopsdef}), we find the inseparability criteria between spatial Stokes operators given by
\begin{eqnarray} \label{stokesinsep}
\mathcal{I}( \hat{S}_{1}, \hat{S}_{2}) & = & \frac{\Delta^2_{x\pm y} \hat{S}_1 + \Delta^2_{x\pm y} \hat{S}_2}{8 |\alpha_{10} \alpha_{01} \sin \theta | }  \nonumber\\
\mathcal{I}( \hat{S}_{3}, \hat{S}_{1}) & = & \frac{\Delta^2_{x\pm y} \hat{S}_1 + \Delta^2_{x\pm y} \hat{S}_3}{8 |\alpha_{10} \alpha_{01} \cos \theta | } \nonumber\\
\mathcal{I}( \hat{S}_{2}, \hat{S}_{3}) & = & \frac{\Delta^2_{x\pm y} \hat{S}_2 + \Delta^2_{x\pm y} \hat{S}_3}{4 |\alpha_{10}^2 - \alpha_{01}^2 | }.
\end{eqnarray}

The correspondence between spatial Stokes and quadrature entanglement becomes obvious when we express the spatial Stokes operator conditional variance $\Delta^2_{x\pm y} \hat{S}_i$ in terms of quadrature operators. Making the assumption that $\alpha_{10} \ll \alpha_{01}$ and using Eq.~(\ref{stokesvar}), gives
\begin{eqnarray}
\Delta_{x\pm y}^2 \hat{S}_{1} & = &  \alpha_{01}^2 \Delta_{x\pm y}^2 \hat{X}_{a_{01}}^+ \nonumber\\
\Delta_{x\pm y}^2 \hat{S}_{2}  & = & \alpha_{01}^2 \Delta_{x\pm y}^2 \hat{X}_{a_{10}}^{\theta} \nonumber\\
\Delta_{x\pm y}^2 \hat{S}_{3} & = &  \alpha_{01}^2 \Delta_{x\pm y}^2  \hat{X}_{a_{10}}^{\theta-\frac{\pi}{2}}. \nonumber\\
\end{eqnarray}
Substituting these expressions into Eq.~(\ref{stokesinsep}), gives
\begin{eqnarray}
\mathcal{I}( \hat{S}_{1}, \hat{S}_{2}) & = & \frac{\alpha_{01}}{8 \alpha_{10} |\sin \theta |} \left( \Delta^2_{x\pm y} \hat{X}_{a_{01}}^+ + \Delta^2_{x\pm y} \hat{X}_{a_{10}}^\theta \right)
\nonumber\\ \label{insep1} \\
\mathcal{I}( \hat{S}_{3}, \hat{S}_{1}) & = & \frac{ \alpha_{01}}{8 \alpha_{10} |\cos \theta |} \left( \Delta^2_{x\pm y} \hat{X}_{a_{01}}^+ + \Delta^2_{x\pm y} \hat{X}_{a_{10}}^{\theta - \frac{\pi}{2}}\right)
\nonumber\\ \label{insep2} \\
\mathcal{I}( \hat{S}_{2}, \hat{S}_{3}) & = & \frac{1}{4}\left( \Delta^2_{x\pm y} \hat{X}_{a_{10}}^\theta + \Delta^2_{x\pm y} \hat{X}_{a_{10}}^{\theta - \frac{\pi}{2}} \right).
\nonumber\\ \label{insep3}
\end{eqnarray}

An inspection of Eqs.~(\ref{insep1}) to (\ref{insep3}) show that in the limit considered here only spatial Stokes entanglement between $\hat{S}_{2}$ and $\hat{S}_{3}$ can be realistically attained. This is because the inseparability criteria of Eqs.~(\ref{insep1}) and (\ref{insep2}) scale with respect to $\alpha_{01}/\alpha_{10}$. Since $\alpha_{01}/\alpha_{10} \gg 1$, the correlation in $\hat{X}_{a_{01}}^+$ and $\hat{X}_{a_{10}}^\theta$ has to be significantly reduced below one, in order to maintain inseparability. On the other hand, we see that after setting $\theta = 0$, Eq.~(\ref{insep3}) reduces to
\begin{equation}
\mathcal{I}( \hat{S}_{2}, \hat{S}_{3}) = \frac{1}{4}\left( \Delta^2_{x\pm y} \hat{X}_{a_{10}}^+ + \Delta^2_{x\pm y} \hat{X}_{a_{10}}^- \right).
\end{equation}
So that quadrature entanglement between the TEM$_{10}$ modes is transformed directly into spatial Stokes entanglement between the $\hat{S}_2$ and $\hat{S}_3$ spatial Stokes operators.

\section{Conclusion}

We have identified the relevant spatial modes for optical beam position, momentum and orbital angular momentum. We formalized the quantum properties of these observables using the Stokes-operator formalism and presented a graphical representation of these variables using the Poincar\'e sphere. A spatial Stokes detection scheme was described and schemes to generate spatial Stokes operator squeezing and entanglement were proposed.

\begin{acknowledgments}

We would like to thank Hans-A.~Bachor for fruitful discussions. This work was supported by the Australian Research Council Centre of Excellence Programme. 

\end{acknowledgments}


\end{document}